\pdfoutput=1
\documentclass[twocolumn,prl,superscriptaddress,longbibliography]{revtex4-2}
\usepackage[colorlinks=true, citecolor=blue, urlcolor=blue, linkcolor=red]{hyperref}
\renewcommand{\section}[1]{{\par\it #1.---}\ignorespaces}
\usepackage{amsmath,amssymb,tikz,scalerel}
\hypersetup{
        colorlinks=true,
        linkcolor=red,
        citecolor=blue,
        urlcolor=blue}
\usetikzlibrary{svg.path}
\definecolor{orcidlogocol}{HTML}{A6CE39}
\tikzset{
	orcidlogo/.pic={
		\fill[orcidlogocol] svg{M256,128c0,70.7-57.3,128-128,128C57.3,256,0,198.7,0,128C0,57.3,57.3,0,128,0C198.7,0,256,57.3,256,128z};
		\fill[white] svg{M86.3,186.2H70.9V79.1h15.4v48.4V186.2z}
		svg{M108.9,79.1h41.6c39.6,0,57,28.3,57,53.6c0,27.5-21.5,53.6-56.8,53.6h-41.8V79.1z M124.3,172.4h24.5c34.9,0,42.9-26.5,42.9-39.7c0-21.5-13.7-39.7-43.7-39.7h-23.7V172.4z}
		svg{M88.7,56.8c0,5.5-4.5,10.1-10.1,10.1c-5.6,0-10.1-4.6-10.1-10.1c0-5.6,4.5-10.1,10.1-10.1C84.2,46.7,88.7,51.3,88.7,56.8z};}}
\newcommand\orcid[1]{\href{https://orcid.org/#1}{\mbox{\scalerel*{\begin{tikzpicture}[yscale=-1,transform shape]\pic{orcidlogo};\end{tikzpicture}}{|}}}}

\begin{document}
\title{Floquet Engineering to Overcome No-Go Theorem of Noisy Quantum Metrology }
\author{Si-Yuan Bai\orcid{0000-0002-4768-6260}}
\affiliation{Key Laboratory of Quantum Theory and Applications of MoE, Lanzhou Center for Theoretical Physics, and Key Laboratory of Theoretical Physics of Gansu Province, Lanzhou University, Lanzhou 730000, China}
\author{Jun-Hong An\orcid{0000-0002-3475-0729}}
\email{anjhong@lzu.edu.cn}
\affiliation{Key Laboratory of Quantum Theory and Applications of MoE, Lanzhou Center for Theoretical Physics, and Key Laboratory of Theoretical Physics of Gansu Province, Lanzhou University, Lanzhou 730000, China}
\begin{abstract}
Permitting a more precise measurement to physical quantities than the classical limit by using quantum resources, quantum metrology holds a promise in developing many revolutionary technologies. However, the noise-induced decoherence forces its superiority to disappear, which is called no-go theorem of noisy quantum metrology and constrains its application. We propose a scheme to overcome the no-go theorem by Floquet engineering. It is found that, by applying a periodic driving on the atoms of the Ramsey spectroscopy, the ultimate sensitivity to measure their frequency characterized by quantum Fisher information returns to the ideal $t^2$ scaling with the encoding time whenever a Floquet bound state is formed by the system consisting of each driven atom and its local noise. Combining with the optimal control, this mechanism also allows us to retrieve the ideal Heisenberg-limit scaling with the atom number $N$.  Our result gives an efficient way to avoid the no-go theorem of noisy quantum metrology and to realize high-precision measurements.
\end{abstract}
\maketitle

\section{Introduction}
Aiming at precise measurements to physical quantities, metrology plays a vital role in developing highly advanced technologies. Restricted by classical statistics, the metrology precision achievable in classical physics is bounded by the shot-noise limit (SNL) $N^{-1/2}$ with $N$ being the number of employed resource in measurements. The advent of quantum metrology reveals that the SNL can be beaten by quantum effects \citep{science.1097576,doi:10.1126/science.1104149,PhysRevLett.96.010401,Giovannetti2011,MA201189,PhysRevLett.107.083601,RevModPhys.89.035002,RevModPhys.90.035005,Daryanoosh2018,PhysRevLett.128.160505,PRXQuantum.3.020310,PRXQuantum.3.010202}. It has been found that, by using entanglement of quantum probes, one can achieve a frequency measurement in Ramsey spectroscopy with a precision of Heisenberg limit (HL) $N^{-1}$.
Quantum metrology has versatile applications in next-generation gyroscope \cite{doi:10.1116/1.5120348,PhysRevApplied.14.034065,PhysRevApplied.14.064023,Jiao:23}, atomic clock \cite{PhysRevLett.111.090801,PhysRevLett.112.190403,Komar2014,PhysRevLett.125.210503,Pedrozo2020}, magnetometers \cite{Thiel2016,Taylor2008,Bao2020}, and gravimetry \cite{PhysRevLett.125.100402}.

The realization of quantum metrology is challenged by the noise-induced decoherence in its stability and scalability \cite{Maccone2011,Escher2011,PhysRevX.12.011039,PhysRevLett.129.240503,PhysRevApplied.17.034073,PhysRevA.102.022618,PhysRevLett.127.060501}. It was found that the Markovian dephasing noise forces the HL in Ramsey spectroscopy not only to return to the SNL at an optimal encoding time but also to become divergent in the long-time condition \cite{PhysRevLett.79.3865}. Being universal for any Markovian noise \cite{Fujiwara_2008,Demkowicz-Dobrzanski2012,PhysRevLett.102.040403,PhysRevA.80.013825,PhysRevLett.92.230801,PhysRevLett.107.113603,Kacprowicz2010}, these two destructive consequences are called no-go theorem of noisy quantum metrology \cite{Albarelli2018restoringheisenberg,PhysRevLett.116.120801}. Further studies showed that the non-Markovian effect of the dephasing noise can reduce the HL to the Zeno limit $N^{-3/4}$ at an optimal time \cite{PhysRevA.84.012103,PhysRevLett.109.233601,PhysRevA.92.010102,PhysRevLett.116.120801,PhysRevLett.123.040402,PhysRevLett.129.070502}. Many efforts, e.g., adaptive \cite{PhysRevLett.111.090801,PhysRevX.7.041009} and nondemolition \cite{PhysRevLett.125.200505} measurements, correlated decoherence \cite{Jeske_2014}, purification \cite{PhysRevLett.129.250503}, error correction \cite{PhysRevLett.112.080801,PhysRevLett.112.150801,PhysRevLett.116.230502,Lu2015,Reiter2017,Zhou2018,PhysRevLett.128.140503}, and dynamical control \cite{PhysRevA.87.032102,Sekatski_2016}, have been proposed to restore the HL. Although partially recovering the quantum superiority in its scaling with $N$, the divergence fate of the precision in the long-time condition does not change. A mechanism to solve the divergence problem was proposed in \cite{Jiao:23,Wang_2017}, but the recovery of the HL for arbitrary $N$ is unavailable. Hence, how to retrieve the HL and overcome the precision divergence with time simultaneously is still an open question.

Inspired by the advance that Floquet engineering has become a versatile tool in quantum control \cite{PhysRevA.91.052122,PhysRevA.102.060201,PhysRevLett.117.250401,PhysRevLett.82.2417} and generating novel quantum phases \cite{Eckardt_2015,McIver2020,Zhang2017}, see a review in Ref. \cite{doi:10.1080/00018732.2015.1055918}, we propose a scheme to overcome the no-go theorem of noisy quantum metrology by Floquet engineering. We discover that, via applying a periodic driving on the atoms of the Ramsey-spectroscopy-based quantum metrology under local dissipative noises, the precision divergence in the long-time condition is avoided as long as a Floquet bound state (FBS) is formed in the quasienergy spectrum of the system formed by each driven atom and its local noise. Further using this mechanism, we can completely recover the HL by optimally designing the driving amplitude. Retrieving the ideal scaling of the precision with $N$ and $t$ simultaneously, our result paves the way to realize quantum metrology in practice.

\begin{figure}
\centering
\includegraphics[width=.85\columnwidth]{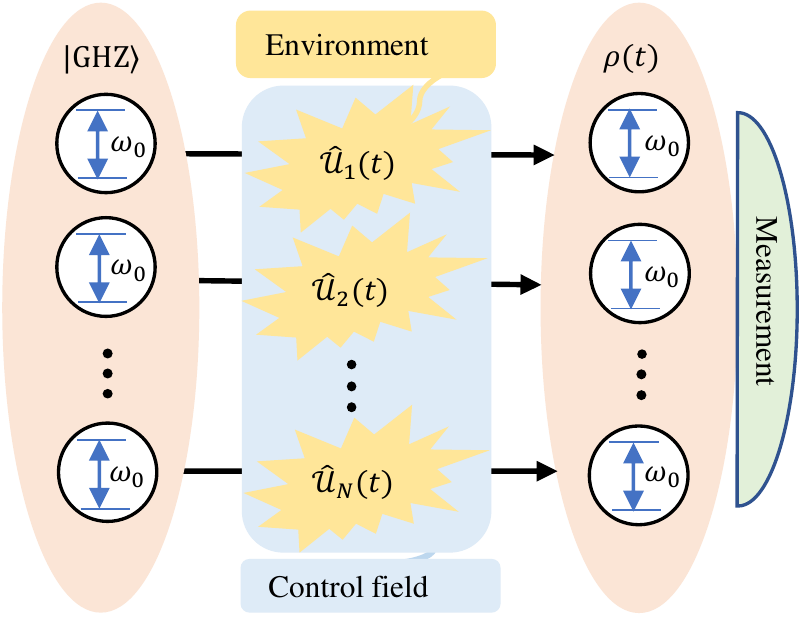}
\caption{Schematic illustration of noisy quantum metrology under periodic control. $\hat{\mathcal{U}}_j(t)$ is the evolution operator of the $j$th periodically driven atom interacting with its environment. }\label{schn}
\end{figure}

\section{Noisy quantum metrology}
Quantum metrology pursues a highly precise measurement of physical quantities by using quantum resources of a probe. To measure a quantity $\theta$ of a system, we first prepare a probe in a state $\rho_\text{in}$ and couple it to the system to encode $\theta$ into the probe state $\rho_\theta$. Then we measure a certain observable $\hat{O}$ of the probe and infer the value of $\theta$ from the result. The inevitable errors make us unable to estimate $\theta$ precisely. According to quantum parameter estimation theory \cite{Liu_2019,PhysRevLett.72.3439}, the ultimate precision of $\theta$ optimizing all possible observable $\hat{O}$ is constrained by the quantum Cram\'{e}r-Rao bound $\delta\theta= 1/\sqrt{\upsilon \mathcal{F}_{\theta}}$, where $\delta\theta$ is the standard error of the estimate, $\upsilon$ is the number of repeated measurements, and $\mathcal{F}_{\theta}=\text{Tr}(\hat{L}_\theta^2\rho_\theta)$ is the quantum Fisher information (QFI) characterizing the most information of $\theta$ extractable from $\rho_\theta$. $\hat{L}_\theta$ called symmetric logarithmic derivative is defined as $\partial_\theta\rho(\theta)=(\hat{L}_\theta\rho_\theta+\rho_\theta \hat{L}_\theta)/2$. If $\delta\theta\propto N^{-1/2}$, with $N$ being the number of the used resource, then the precision is called the SNL, which is the achievable limit of any classical measurement. The SNL can be beaten by using quantum protocols.
\begin{figure*}
\includegraphics[width=1\textwidth]{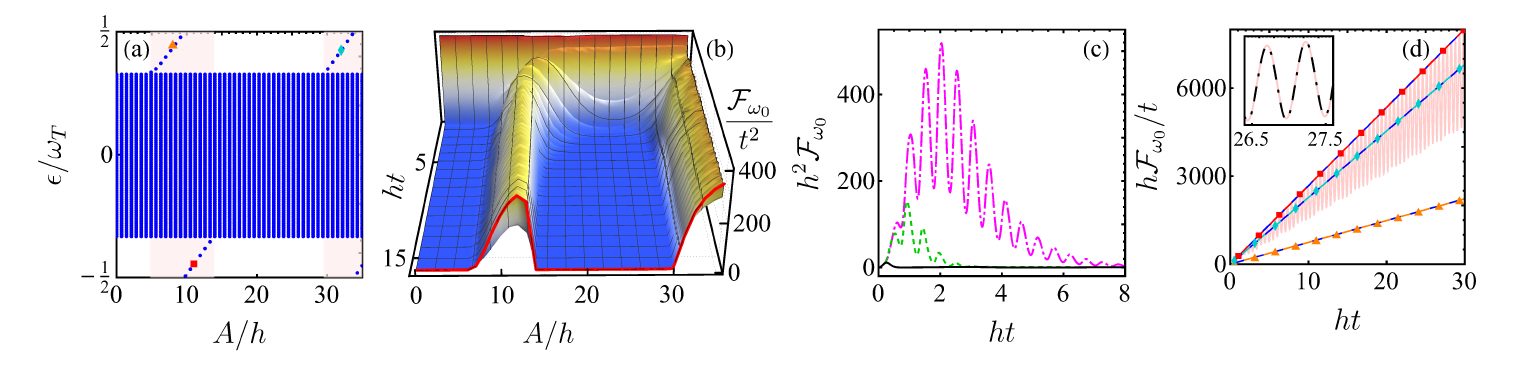}
\caption{(a) Quasienergy spectrum in different driving amplitude $A$. The shaded backgrounds mark the regions with the FBS. (b) Evolution of $\mathcal{F}_{\omega_{0}}/t^{2}$ in different $A$ via numerically solving Eq. \eqref{eq:dynamics}. The red line shows the result evaluated from Eq. \eqref{fud} at $t=31T$. (c) Evolution of $\mathcal{F}_{\omega_0}$ in the absence of the FBS when $A=16h$ (magenta dot-dashed line), $20h$ (green dashed line), and $0$ (black line). (d) Evolution of $\mathcal{F}_{\omega_0}/t$ in the presence of the FBS when $A=11h$ (light-red line), whose stroboscopic values are given by the red line marked by the red squares. The lines marked by the orange triangle and the cyan diamond are the stroboscopic values of $\mathcal{F}_{\omega_{0}}/t$ when $A=8h$ and $32h$, respectively. The blue dashed lines are the QFI evaluated from Eq. \eqref{lotstd}, which match with the stroboscopic values in the long-time limit. The insert is the comparison of the numerical result (light-red line) with the one evaluated from the FBS (black dot-dashed line). We use $\omega_{T}=12h$, $g=\omega_0=h$, $\omega_c=0$, and $N=20$. }\label{fig:Fig-1}
\end{figure*}

In a Ramsey spectroscopy to measure the atomic frequency $\omega_0$ \cite{PhysRevLett.79.3865}, one chooses $N$ atoms themselves as the probe and prepares their state in a Greenberger-Horne-Zeilinger (GHZ) type entangled state $|\psi_\text{in}\rangle=(|g\rangle^{\otimes N}+|e\rangle^{\otimes N})/\sqrt{2}$, where $|g\rangle$ and $|e\rangle$ are the atomic ground and excited states. Then the free evolution governed by the $j$th atomic Hamiltonian $\hat{H}_{0,j}=\omega_0\hat{\sigma}_j^\dag\hat{\sigma}_j$, with $\hbar=1$ and $\hat{\sigma}_j=|g_j\rangle\langle e_j|$, encodes $\omega_0$ into the state $|\psi(t)\rangle=e^{-i\sum_j\hat{H}_{0,j}t}|\psi_\text{in}\rangle$. The ultimate precision described by the QFI is evaluated as $\mathcal{F}_{\omega_0}(t)=N^2t^2$, which scales with $N$ as the HL and beats the classical SNL $N$ times due to the utilized entanglement \cite{PhysRevLett.112.190403,doi:10.1126/science.1097576,PhysRevLett.102.100401}. It exhibits the quantum superiority of the metrology scheme. Another feature is that $\mathcal{F}_{\omega_0}(t)$ scales with the encoding time as $t^2$, which also acts as a resource to increase the precision \cite{Pang2017}.

In practice, the encoding process is influenced by the dissipative environments, which cause the energy loss and the entanglement degradation of the probe. The encoding is governed by $\hat{H}=\sum_j\hat{H}_{j}$ with
\begin{equation}
\hat{H}_j=\hat{H}_{0,j}+\sum_{k}[\omega_k\hat{a}_{j,k}^\dag\hat{a}_{j,k}+g_{j,k}(\hat{a}_{j,k}\hat{\sigma}_j^\dag+\text{H.c.})],
\end{equation}
where $\hat{a}_{j,k}$ is the annihilation operator of the $k$th mode with frequency $\omega_k$ of the environment felt by the $j$th atom and $g_{j,k}$ is their coupling strength. The coupling is further described by the spectral density $J(\omega)=\sum_k|g_k|^2\delta(\omega-\omega_k)$. After tracing out the environmental degrees of freedom from the dynamics of the total system, we obtain the non-Markovian master equation \cite{Wang_2017}
\begin{equation}
\dot{\rho}(t)=\sum_{j=1}^N\{-i\omega(t)[\hat{\sigma}_j^\dag\hat{\sigma}_j,\rho(t)]+\gamma(t)\check{\mathcal{L}}_j\rho(t)\},\label{maseq}
\end{equation}
where $\check{\mathcal{L}}_{j}\rho(t)\equiv 2\hat{\sigma}_{j}\rho(t)\hat{\sigma}_{j}^{\dag}-\{\rho(t),\hat{\sigma}_{j}^{\dag}\hat{\sigma}_{j}\}$, $\gamma(t)\equiv-\mathrm{Re}[\dot{c}(t)/c(t)]$ is the dissipation rate, and $\omega(t)\equiv-\mathrm{Im}[\dot{c}(t)/c(t)]$ is the renormalized frequency. Here, $c(t)$ is determined by the integro-differential equation
\begin{equation}~\label{ct}
\dot{c}(t)+i\omega_{0}c(t)+\int_{0}^{t}\nu(t-\tau)c(\tau)d\tau=0,
\end{equation}
where $\nu(x)\equiv\int_{0}^{\infty}J(\omega)e^{-i\omega x}d\omega$ is the environmental correlation function and $c(0)=1$. Solving Eq. \eqref{maseq} under the initial state $|\psi_\text{in}\rangle$, we can calculate the corresponding QFI according to \citep{Liu_2016,Liu_2019}
\begin{equation}
\mathcal{F}_{\omega_0}=\sum_{i, j}[\lambda^{-1}_{i}{\lambda}_{i}^{\prime2}+4\lambda_{i}\langle\lambda'_{i}|\lambda'_{i}\rangle-
\frac{8\lambda_{i}\lambda_{j}}{\lambda_{i}+\lambda_{j}}|\langle\lambda_{i}|\lambda'_{j}\rangle|^{2}],~~~\label{ghzf1}
\end{equation}
where the prime denotes the derivative with respect to $\omega_0$ and $\rho(t)=\sum_i\lambda_i|\lambda_i\rangle\langle\lambda_i|$.

In the special case when the probe-environment coupling is weak and the characteristic timescale of the environment is smaller than that of the probe, we can apply the Markovian approximation to Eq.~(\ref{ct}). The approximate solution reads
$c_\text{MA}(t)= \exp\{-\kappa t-i[\omega_{0}+\Delta(\omega_{0})]t\}$~\cite{PhysRevE.90.022122,SupplementalMaterial}, where $\kappa=\pi J(\omega_{0})$ and $\Delta(\omega_{0})=\mathcal{P}\int_{0}^{\infty}\frac{J(\omega)}{\omega_{0}-\omega}d\omega$. Here, $\mathcal{P}$ denotes the Cauchy principal value. The QFI is $\mathcal{F}^\text{MA}_{\omega_0}(t)=\frac{2 N^{2}t^{2} }{1+(e^{2\kappa t}-1)^{N}+e^{2 N \kappa t}}$. First, we find that $\mathcal{F}^\text{MA}_{\omega_0}(t)$ has a maximum $\max_{t}\mathcal{F}^\text{MA}_{\omega_{0}}=0.24/\kappa^{2}$ when $t_\text{opt}=1.11/(\kappa N)$ in the large-$N$ condition. In the Ramsey spectroscopy, one generally repeats the measurement within a time duration $T_R$. Thus, we have the repeating times $\upsilon=T_R/t_\text{opt}$. According to quantum Cram\'{e}r-Rao bound, the ultimate precision is $\min_t\delta\omega^\text{MA}_0= (0.22T_R N/\kappa)^{-1/2}$, which is the SNL \cite{PhysRevLett.79.3865}. Second, we see $\lim_{t\rightarrow \infty}\mathcal{F}^\text{MA}_{\omega_0}(t)=0$ and $\lim_{t\rightarrow \infty}\delta\omega_0^\text{MA}(t)=\infty$, which implies the breakdown of the scheme in the long-time condition. Thus, the noise washes out the quantum superiority. This is called the no-go theorem of noisy quantum metrology \cite{PhysRevLett.116.120801,Albarelli2018restoringheisenberg}.

\section{Floquet engineering}
To overcome the no-go theorem, we apply a periodic driving $\hat{H}_{c,j}(t)=f(t)\hat{\sigma}^\dag_j\hat{\sigma}_j$, with $f(t)={A\over 2}[1-\cos(\omega_Tt)]$ and $\omega_T$ being the driving frequency, on the probe. This is realizable by a time-dependent Zeeman magnetic field \cite{PhysRevA.105.043112,Hartung2019,PhysRevLett.113.263005,PhysRevLett.118.163203}. It is interesting to find that the master equation in this case takes the same form as Eq. \eqref{maseq}, but Eq. \eqref{ct} becomes
\begin{equation}
\dot{c}(t)+i[\omega_{0}+f(t)]c(t)+\int_{0}^{t}d\tau \nu(t-\tau)c(\tau)=0.\label{eq:dynamics}
\end{equation}
Because of the dynamical independence of each atom, Eq. \eqref{eq:dynamics} is equivalent to $c(t)=\langle e_j,\{0_k\}|\hat{\mathcal{U}}_j(t)|e_j,\{0_k\}\rangle$ for each atom and its local environment, where $\hat{\mathcal{U}}_j(t)=\hat{\mathcal{T}}e^{-i\int_0^t[\hat{H}_j+\hat{H}_{c,j}(\tau)]d\tau}$ is the evolution operator, $|\{0_k\}\rangle$ is the environmental vacuum state, and $\hat{\mathcal{T}}$ is the time-ordering operator. According to the Floquet theorem, $\hat{\mathcal{U}}_j(t)$ for the periodic system $\hat{H}_j(t)=\hat{H}_j(t+T)$, with $T=2\pi/\omega_T$, is expanded as $\hat{\mathcal{U}}_j(t)=\sum_\alpha e^{-i\epsilon_\alpha t}|u_\alpha(t)\rangle\langle u_\alpha(0)|$. Here $\epsilon_\alpha$ and $|u_\alpha(t)\rangle=|u_\alpha(t+T)\rangle$, being called quasienergies and quasistationary states, respectively, are determined by the Floquet eigen equation $[\hat{H}_j(t)-i\partial_t]|u_\alpha(t)\rangle=\epsilon_\alpha|u_\alpha(t)\rangle$ \cite{PhysRev.138.B979,PhysRevA.7.2203}. The quasienergy spectrum of a periodically driven two-level system coupled to an environment constitutes a finite-width quasienergy band and possibly formed discrete quasienergy levels, which we call FBSs. The contribution of the quasistationary states from the quasienergy band to the long-time $c(t)$ tends to zero due to the out-of-phase interference \cite{PhysRevA.91.052122,PhysRevA.102.060201}. Only the ones from FBSs survive. Then the long-time solution of Eq. \eqref{eq:dynamics} under the stroboscopic dynamics, i.e., $t=nT$, is \cite{SupplementalMaterial}
\begin{eqnarray}
\lim_{t\rightarrow\infty}c(t)=\sum_{l=1 }^MZ_{l}e^{-i\epsilon^\text{b}_{l}t}\label{lotstd}
\end{eqnarray}
when $M$ FBSs are formed, where $Z_{l}=|\langle u^\text{b}_{l}(0)|e_j,\{0_{k}\}\rangle |^2 $ and $|u^\text{b}_{l}(t)\rangle$ is the $l$th FBS with quasienergy $\epsilon^\text{b}_{l}$.

In the absence of the FBS, it is natural to expect that $\mathcal{F}_{\omega_0}(t)$ tends to zero because $c(t)$ asymptotically approaches zero. Consistent with the Markovian result, the metrology scheme in this case is also broken by the dissipation. We focus on the case that one FBS is formed. Substituting Eq. \eqref{lotstd} into Eqs. \eqref{maseq} and \eqref{ghzf1} under the initial GHZ state, we obtain \cite{SupplementalMaterial}
\begin{equation}
\lim_{N,t\rightarrow\infty}\mathcal{F}_{\omega_0}(t)\simeq y_N(Z^2)(Nt\partial_{\omega_0}\epsilon^\text{b})^2+{N(\partial_{\omega_0}Z^2)^2\over Z^2(1-Z^2)},\label{fud}
\end{equation}
where $y_N(x)=2x^N/[1+x^N+(1-x)^N]$. We find that Eq. \eqref{fud} returns to its ideal $t^2$ scaling with the encoding time up to a time-independent shift of the second term. Furthermore, its scaling relation with $N$ would return to the HL if $y_N(Z^2)$ is managed to be recast into a $N$-independent function. It is realizable by assuming $\text{opt}Z^2=e^{-a/N}$, which results in $y_N(Z^2)\backsimeq\frac{2}{e^{a}+1}$ and
\begin{equation}\text{opt}_A\lim_{N,t\rightarrow\infty}\mathcal{F}_{\omega_0}(t) = \frac{2}{e^{a}+1}(N t \partial_{\omega_0} \epsilon^\text{b})^2 ,\label{optfi}
\end{equation}
with $a$ being a positive constant. Then we use $\text{opt}Z^2$ to inversely design the optimal driving amplitude $A$ for given $N$. Thus, it is remarkable to find that the ideal $(Nt)^2$ scaling of $\mathcal{F}_{\omega_0}(t)$ is recovered and the no-go theorem is overcome via Floquet engineering and optimal control. Note that our result goes beyond the ones in \cite{Jiao:23,Wang_2017}, where, although the error divergence with time was avoided, the sensitivity was even worse than the SNL for large $N$.

\begin{figure*}
\includegraphics[width=1\textwidth]{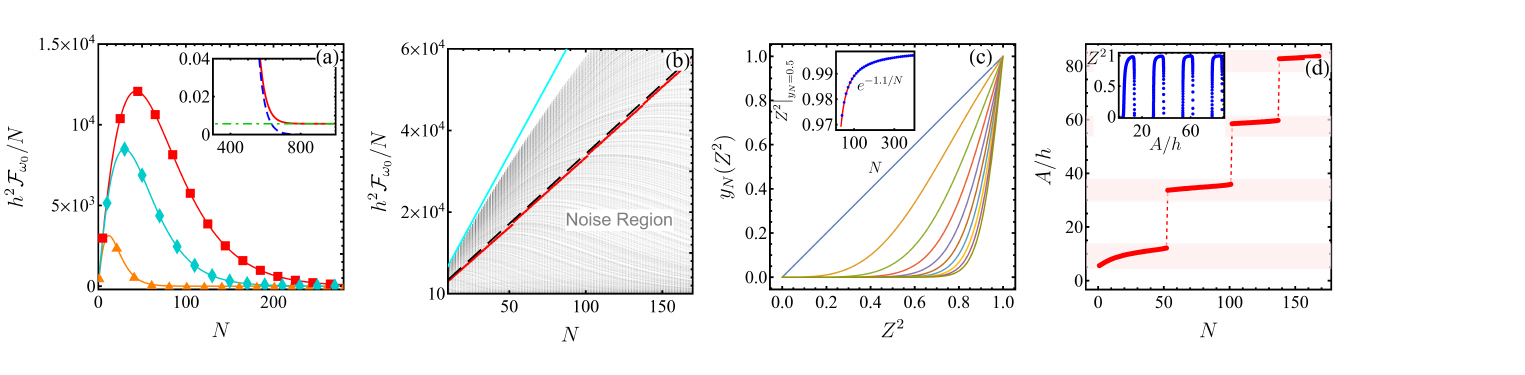}
\caption{Stroboscopic values of $\mathcal{F}_{\omega_{0}}/N$ at $t=50T$ as a function of the atom number $N$ in (a) for $A=8h$ (the orange triangle line), $11h$ (the red square line), and $32h$ (the cyan diamond line), and in (b) for all $A\in[0,100]h$ supporting the FBS (the gray lines). The inset of (a) shows the large-$N$ behavior for $A=32h$, where the blue dashed and green dot-dashed lines are from the first and second terms of Eq. \eqref{fud}, respectively. The cyan in (b) is the ideal HL. The red line is the numerical result under the optimized $A$ such that $\text{opt}_{A} y_{N}(Z^{2})=0.5$.  The black dashed line is the analytic result evaluated from Eq. \eqref{fud}. (c) $f_{N}(Z^{2})$ as a function of $Z^{2}$ for $N$ ranging from $1$ to $19$ with a fixed interval $2$ by the lines from left to the right. The dependence of $Z^2$ on $N$ to make $y_{N}(Z^{2})=0.5$ is denoted by the blue dots in the inset, which are numerically fitted as $Z^{2}=e^{-1.1/N}$. (d) Optimized $A$ in different $N$ such that $y_{N}(Z^{2})=0.5$, which causes $\lim_{t\rightarrow\infty}\mathcal{F}_{\omega_0}\propto N^2$ shown by the red line in (b). The inset shows $Z^2$ in different $A$. The shaded backgrounds mark the regions where the FBS is formed. Other parameters are the same as Fig. \ref{fig:Fig-1}.}\label{opmtzed}\end{figure*}

\section{Numerical results}
We consider that the environment consists of a two-dimensional structured reservoir, with a finite bandwidth $8h$, whose dispersion relation reads $\omega_{\bf k}=\omega_c-2h(\cos k_{x}+\cos k_{y})$ \cite{PhysRevLett.119.143602,PhysRevLett.122.203603}. It may be realized by a $L\times L$ coupled resonator array or optical lattice with $\omega_c$ being the eigen frequency of the resonator or the lattice and $h$ being the nearest-neighbor hopping rate. The atom-environment coupling strength is $g_{j,{\bf k}}=ge^{-i{\bf k}\cdot{\bf r}_j}/L$. A prerequisite for forming the FBS is the existence of band gaps in the quasienergy spectrum, which occurs when $\omega_T>8h$ \cite{SupplementalMaterial}. The quasienergy spectrum of the total system formed by each driven atom and its environment in Fig. \ref{fig:Fig-1}(a) shows that, with increasing the driving amplitude $A$, the two branches of FBS quasienergies residing in the band gaps divide the spectrum into two regions: without the FBS when $A\in(0,4.9)h\cup (13.9,28.3)h$ and one FBS when $A\in(4.9,13.9)h\cup(28.3,35.2)h$. Via numerically solving Eq. \eqref{eq:dynamics}, we obtain the evolution of $\mathcal{F}_{\omega_{0}}/t^{2}$ in different $A$ [see Fig. \ref{fig:Fig-1}(b)]. In the region without the FBS, $\mathcal{F}_{\omega_{0}}/t^{2}$ exclusively decays to zero, which exhibits the destructive effect of the no-go theorem. It is interesting to observe that, as long as the FBS is formed, $\mathcal{F}_{\omega_{0}}/t^{2}$ tends to a nonzero constant with a persistent tiny-amplitude oscillation. The red line in Fig. \ref{fig:Fig-1}(b) from Eq. \eqref{fud} indicates that the peaks of the oscillation match with the stroboscopic dynamics. This confirms that the $t^{2}$ scaling of $\mathcal{F}_{\omega_0}$ in the ideal case is recovered in the long-time condition. A detailed analysis of the FBS-absence cases in Fig. \ref{fig:Fig-1}(c) shows that, although being dramatically enhanced over the one without the driving (black solid line), $\mathcal{F}_{\omega_0}$ under the driving decays to zero finally after a transient increasing. On the contrary, the stroboscopic values of $\mathcal{F}_{\omega_0}/t$ in the presence of the FBS when $A=11h$ [see the light-red line in Fig. \ref{fig:Fig-1}(d)], which match the analytical results in Eq. \eqref{lotstd}, show a linear increasing with $t$. This confirms the $t^2$ scaling of $\mathcal{F}_{\omega_0}$ in the long-time dynamics. The similar behavior is observed for $A=8h$ and $32h$ in Fig. \ref{fig:Fig-1}(d). The matching of the numerical result with the one from the FBS in the inset of Fig. \ref{fig:Fig-1}(d) verifies the dominant role of the FBS in the long-time dynamics. All these results indicate that the problem of the error divergence with time has been avoided due to the formation of the FBS.

To reveal whether the HL is retrievable or not when the FBS is formed, we plot in Fig. \ref{opmtzed}(a) the long-stroboscopic time $\mathcal{F}_{\omega_{0}}/N$ as a function of atom number $N$. It indicates a near-linear increasing of $\mathcal{F}_{\omega_{0}}/N$ with $N$, i.e., $\mathcal{F}_{\omega_{0}}\propto N^2$, only for small $N$. For large $N$, $\mathcal{F}_{\omega_{0}}/N$ exhibits an unwanted exponential decay to a constant governed by the second term of Eq. \eqref{fud} [see the inset of Fig. \ref{opmtzed}(a)]. Being similar to the result in the absence of the periodic driving \cite{Jiao:23,Wang_2017}, such a $N$-dependent behavior is universal to all the driving amplitude $A$ supporting the FBS [see Fig. \ref{opmtzed}(b)]. We find that $\mathcal{F}_{\omega_0}/N$ in different $A$ fill up the region below the ideal HL. This implies the possibility to avoid the decay of $\mathcal{F}_{\omega_{0}}$ with $N$ and thus to retrieve the $N^{2}$ scaling of $\mathcal{F}_{\omega_{0}}$ by optimizing $A$. For this purpose, we plot in Fig. \ref{opmtzed}(c) $y_N(Z^2)$ as a function of $Z^2$ in different $N$. It exponentially decreases with increasing $N$ for a definite $Z^2$. Because all the curves of $y_N(Z^2)$ for different $N$ have two same endpoints, $(0,0)$ and $(1,1)$, we can always find an optimal $Z^2$ for a given $N$ to make $y_N(Z^2)$ being a constant. By requesting $y_{N}(Z^{2})=0.5$, we plot in the inset of Fig. \ref{opmtzed}(c) the needed $Z^{2}$ as a function of $N$. The numerical fitting reveals $\text{opt}Z^2|_{y_{N}(Z^{2})=0.5}=e^{-1.1/N}$, which permits us to design the optimal $A$ from the numerical correspondence between $Z^2$ and $A$ in the inset of Fig. \ref{opmtzed}(d). Figure \ref{opmtzed}(d) shows the obtained optimal $A$ in different $N$. The corresponding $\text{opt}\lim_{t\rightarrow\infty}\mathcal{F}_{\omega_0}(t)$ is plotted by the red line in Fig. \ref{opmtzed}(b), which matches with our analytical result in Eq. \eqref{optfi}. It clearly depicts that the decay with $N$ is avoided and the HL $\mathcal{F}_{\omega_{0}}\propto N^2$ is retrieved by the optimally designed periodic driving.  The results reveal that, given an atom number $N$, we always have an experimentally realizable optimal periodic driving such that the no-go theorem of noisy quantum metrology is completely overcome.

\section{Discussion and conclusion}
Going beyond the widely used Markovian approximation \cite{PhysRevLett.111.090801,PhysRevX.7.041009,PhysRevLett.125.200505,Jeske_2014,PhysRevLett.129.250503,PhysRevLett.112.080801,PhysRevLett.112.150801,PhysRevLett.116.230502,Lu2015,Reiter2017,Zhou2018,PhysRevLett.128.140503}, our scheme is an exact characterization to remove the noise effects on quantum metrology via periodic driving. This enables us retrieve the ideal HL scaling of the precision in the Ramsey spectroscopy with respect to not only the atom number but also the encoding time. It succeeds in overcoming the challenge set by decoherence on the scalability and stability of quantum metrology.  Although only the GHZ state is studied, our FBS mechanism is applicable to other initial states, e.g., the spin squeezing state \cite{MA201189,PhysRevLett.127.083602}. Because of the building block role of Ramsey spectroscopy in developing quantum gyroscope, atomic clock, magnetometers, and gravimetry \cite{doi:10.1116/1.5120348,PhysRevApplied.14.034065,PhysRevApplied.14.064023,Jiao:23,PhysRevLett.111.090801,PhysRevLett.112.190403,Komar2014,PhysRevLett.125.210503,Pedrozo2020,Thiel2016,Taylor2008,Bao2020,PhysRevLett.125.100402}, our result has a practical meaning in realizing these technique innovations. Floquet engineering is widely used in quantum technologies \cite{PhysRevA.91.052122,PhysRevA.102.060201,PhysRevLett.117.250401,PhysRevLett.82.2417,Eckardt_2015,McIver2020,Zhang2017,doi:10.1080/00018732.2015.1055918}. Without resorting to the well designed control pulses \cite{PhysRevA.87.032102,Sekatski_2016,PhysRevA.96.012117,PhysRevA.96.042114}, our always-on sinusoidal periodic driving meets the experimental accessibility. The system-environment bound state in the static case has been observed in circuit QED \cite{Liu2017} and cold atom \cite{Krinner2018} systems, which provides a support to realize our FBS in periodically driven systems.

In summary, we have proposed a Floquet engineering scheme to eliminate the detrimental effects of dissipative noises on quantum metrology. Via applying a periodic driving on the atoms of the Ramsey-spectroscopy-based quantum metrology in the presence of local noises, we have found that its precision characterized by the QFI is essentially determined by the quasienergy-spectrum feature of the system formed by each driven atom and its noise. Whenever a FBS is formed in the spectrum, the ideal $t^2$ scaling of the QFI with the encoding time is recovered. Further combining the optimal control to the driving field, the HL scaling of the QFI with arbitrary atom number $N$ is also retrieved. Giving a flexible way to solve the long-standing no-go theorem of noisy quantum metrology, our scheme supplies a guideline to realize the ultrasensitive measurement in practical situation.

\section{Acknowledgments}
The work is supported by the National Natural Science Foundation (Grants No. 12205128, No. 12275109, No. 12247101, and No. 11834005), China Postdoctoral Science Foundation (Grants No. BX20220138 and No. 2022M710063), and the Supercomputing Center of Lanzhou University.

\bibliography{FM}
\end{document}


\title{Supplemental Material for ``Floquet engineering to overcome no-go theorem of noisy quantum metrology'' }
\author{Si-Yuan Bai\orcid{0000-0002-4768-6260}}
\affiliation{Key Laboratory of Quantum Theory and Applications of MoE, Lanzhou Center for Theoretical Physics, and Key Laboratory of Theoretical Physics of Gansu Province, Lanzhou University, Lanzhou 730000, China}
\author{Jun-Hong An\orcid{0000-0002-3475-0729}}
\email{anjhong@lzu.edu.cn}
\affiliation{Key Laboratory of Quantum Theory and Applications of MoE, Lanzhou Center for Theoretical Physics, and Key Laboratory of Theoretical Physics of Gansu Province, Lanzhou University, Lanzhou 730000, China}
\maketitle

\section{No-go theorem of noisy quantum metrology}
The solution of Eq. (2) in the main text for the initial GHZ state reads
\begin{eqnarray}~\label{app:app5}
  \rho(t)&=&\frac{1}{2}\Big{[}p_{t}|e\rangle\langle e|+(1-p_{t})|g\rangle\langle g|\Big{]}^{\otimes N}\nonumber\\
  &&+\frac{1}{2}\Big{[}|g\rangle\langle g|^{\otimes N}+\Big{(}c(t)^{N}|e\rangle\langle g|^{\otimes N}+\mathrm{H}.\mathrm{c}.\Big{)}\Big{]}.
\end{eqnarray}
Expanding the first term of Eq. \eqref{app:app5} as
\begin{eqnarray}~\label{app:app6}
&&\big{[}p_{t}|e\rangle\langle e|+(1-p_{t})|g\rangle\langle g|\big{]}^{\otimes N}\nonumber\\
&=&p_{t}^{N}|e\rangle\langle e|^{\otimes N}+\big{(}1-p_{t}\big{)}^{N}|g\rangle\langle g|^{\otimes N}+\sum_{\mathbb{P}}\sum_{m=1}^{N-1}p_{t}^{m}\nonumber\\
&&\times\big{(}1-p_{t}\big{)}^{N-m}
\mathbb{P}\big{[}|e\rangle\langle e|^{\otimes m}\otimes|g\rangle\langle g|^{\otimes(N-m)}\big{]},
\end{eqnarray}
where $\mathbb{P}$ represents all bipartite permutations and $p_t=|c(t)|^2$, we rearrange Eq. \eqref{app:app5} as $\rho(t)=\rho_{1}(t)\oplus\rho_{2}(t)$ with
\begin{eqnarray}
\rho_{1}(t)&=&\frac{1}{2}\big{\{}p_{t}^{N}|e\rangle\langle e|^{\otimes N}+\big{[}1+(1-p_{t})^{N}\big{]}|g\rangle\langle g|^{\otimes N}\nonumber\\&&+\big{[}c(t)^{N}|e\rangle\langle g|^{\otimes N}+\mathrm{H}.\mathrm{c}.\big{]}\big{\}},\label{app:app7}\\
\rho_{2}(t)&=&\frac{1}{2}\sum_{\mathbb{P}}\sum_{m=1}^{N-1}p_{t}^{m}\big{(}1-p_{t}\big{)}^{N-m}\mathbb{P}\big{[}|e\rangle\langle e|^{\otimes m}\nonumber\\&&\otimes|g\rangle\langle g|^{\otimes(N-m)}\big{]}.\label{app:app8}
\end{eqnarray}
It is easy to see that $\rho_{1}(t)$ is a two-by-two matrix in the basis $\{|e\rangle^{\otimes N},|g\rangle^{\otimes N}\}$ and $\rho_{2}(t)$ is a $(2^{N}-2)$-dimensional diagonal matrix. According to the fact that the quantum Fisher information (QFI) is additive for a direct-sum density matrix \cite{Liu_2016}, we have $\mathcal{F}_{\omega_0}(t)=\mathcal{F}_{\omega_0}^{(1)}+\mathcal{F}_{\omega_0}^{(2)}$, with $\mathcal{F}_{\omega_0}^{(1,2)}$ being the QFI of $\rho_{1,2}(t)$.

The QFI for an arbitrary state $\rho$ reads \cite{Liu_2019}
\begin{equation}~\label{app:app9}
\mathcal{F}_{\omega_0}=\sum_{i,j=1}^{D}\bigg{[}\frac{{\lambda'}_{i}^{2}}{\lambda_{i}}+4\lambda_{i}\langle\lambda'_{i}|\lambda'_{i}\rangle-\frac{8\lambda_{i}\lambda_{j}}{\lambda_{i}+\lambda_{j}}|\langle\lambda_{i}|\lambda'_{j}\rangle|^{2}\bigg{]},
\end{equation}where $\rho=\sum_{i=1}^{M}\lambda_{i}|\lambda_{i}\rangle\langle\lambda_{i}|$ and $D$ is the number of the nonzero eigenvalues of $\rho(t)$. Since the eigenvectors of the diagonal matrix $\rho_{2}(t)$ are independent of $\omega_0$, the nonzero component is only the first term of Eq. \eqref{app:app9}
\begin{eqnarray}~\label{app:app10}
  \mathcal{F}_{\omega_0}^{(2)}&=&\frac{1}{2}\sum_{m=1}^{N-1}\frac{N!}{m!(N-m)!}\frac{\{[p_{t}^{m}(1-p_{t})^{N-m}]'\}^{2}}{p_{t}^{m}(1-p_{t})^{N-m}}\nonumber\\
  &=&-\frac{Np_{t}^{\prime2}}{2 p_{t}}\frac{}{}\bigg{[}\frac{Np_{t}(1-p_{t})^{N}+p_{t}-1}{(p_{t}-1)^{2}}+Np_{t}^{N-1}\bigg{]}.~~~~~~
\end{eqnarray}
 With Eq. \eqref{app:app9} and the eigendecomposition of $\rho_{1}(t)$ at hand, $\mathcal{F}_{\omega_0}^{(1)}$ is readily calculated.

The Markovian approximate solution of Eq. (3) in the main text reads $c_\text{MA}(t)=\exp\{-\kappa t-i[\omega_{0}+\Delta(\omega_{0})]t\}$~\cite{PhysRevE.90.022122}, where $\kappa=\pi J(\omega_{0})$ is the decay rate and $\Delta(\omega_{0})=\mathcal{P}\int_{0}^{\infty}\frac{J(\omega)}{\omega_{0}-\omega}d\omega$ is the renormalized frequency. Here, $\mathcal{P}$ denotes the Cauchy principal value. Treating the decay rate $\kappa$ constant, we readily have $p_t=e^{-2\kappa t} $ being independent of $\omega_0$, which results in $\mathcal{F}_{\omega_0}^{(2)}=0$. Substituting $c_\text{MA}(t)$ into Eq. \eqref{app:app7} and diagonalizing the two-by-two matrix $\rho_1(t)$, we straightforwardly obtain
\begin{equation}
	\mathcal{F}_{\omega_0}(t)=	\mathcal{F}_{\omega_0}^{(1)}=\frac{2 N^{2}t^{2}}{1+(e^{2\kappa t}-1)^{N}+ e^{2 N \kappa t}}.\label{smfqf}
\end{equation}
First, Eq. \eqref{smfqf} reduces to the ideal result $\mathcal{F}^\text{ideal}_{\omega_0}(t)=N^2t^2$ when $\kappa=0$. Second, Eq. \eqref{smfqf} reveals
\begin{equation}
\lim_{t\rightarrow\infty}\mathcal{F}_{\omega_0}(t)=0,
\end{equation}
which, in sharp constrast to the ideal case, means that the superiority of the encoding time $t$ as a resource in enhancing the precision is destroyed by the noise. Third, a maximal value $\mathcal{F}_{\omega_0}(t)$ is obtained as
\begin{equation}
\max_{t}\mathcal{F}_{\omega_0}(t)=\frac{2q^{2}}{1+(e^{\frac{2\kappa q}{N}}-1)^{N}+e^{2q\kappa}}\label{smmaq}
\end{equation}
at an optimal time $t_\text{opt}=q/N$, where $q=\frac{1}{2\kappa}[W\left(\frac{2}{e^{2}}\right)+2]=1.11/\kappa$ and $W(z)$ is the Lambert $W$ function. In the large $N$ limit, Eq. \eqref{smmaq} reduces into
\begin{equation}
	\lim_{N\rightarrow\infty}\max_{t}\mathcal{F}_{\omega_0}(t)=\frac{2q^{2}}{1+e^{2q\kappa}}={0.24\over\kappa^2}.
\end{equation}
In the Ramsey spectroscopy, one repeats the measurement in a time duration $T_R$. Thus, the repeating times read $\upsilon=T_R/t_\text{opt}=T_R N/q$. According to the quantum Cram\'{e}r-Rao bound, the ultimate precision of $\omega_0$ is
\begin{equation}
\min\delta\omega_0=\Big[\upsilon \lim_{N\rightarrow\infty}\max_{t}\mathcal{F}_{\omega_0}(t)\Big]^{-1/2}=(0.22T_R N/\kappa)^{-1/2},
\end{equation}
which returns to the shot-noise limit. Therefore, the Markovian noise forces the ideal Heisenberg-limit precision in Ramsey spectroscopy not only to return to the shot-noise limit at an optimal encoding time but also to become
divergent in the long-time condition. The superiority of the encoding time $t$ and the atom number $N$ as resources in enhancing the precision is completely destroyed by the noise. This is called the no-go theorem of noisy quantum metrology.

\section{Steady-state solution of $c(t)$}
We consider first the local dynamics of the periodically driven $j$th atom coupled to its environment governed by $\hat{H}_j(t)=\hat{H}_{j}+\hat{H}_{c,j}(t)$ under the initial condition $|\psi(0)\rangle=|e_j,\{0_k\}\rangle$. Its evolved state $|\psi(t)\rangle=\hat{\mathcal{U}}_j(t)|\psi(0)\rangle$ is expanded as
\begin{equation}
|\psi(t)\rangle=[c(t)\hat{\sigma}_j^\dag+\sum_k d_k(t)\hat{a}_{k,j}^\dag]|g_j,\{0_k\}\rangle,
\end{equation}where $|\{0_k\}\rangle$ is the environmental vacuum state, $c(t)$ is the probability amplitude of the $j$th atom being in the excited state and the environment being in the vacuum state, and $d_k(t)$ is the probability amplitude of the $j$th atom being in the ground state and the environment containing only one photon in the $k$th mode. One can derive from Schr\"{o}dinger equation $i|\dot{\psi}(t)\rangle=\hat{H}_j(t)|\psi(t)\rangle$ that the probability amplitude $c(t)$ satisfies
\begin{equation}
\dot{c}(t)+i[\omega_{0}+f(t)]c(t)+\int_{0}^{t}d\tau \nu(t-\tau)c(\tau)=0,\label{smdynamics}
\end{equation}
which is just Eq. (5) in the main text. Therefore, the solution of Eq. (5) is equivalent to
\begin{equation}
c(t)=\langle e_j,\{0_k\}|\psi(t)\rangle=\langle e_j,\{0_k\}|\hat{\mathcal{U}}_j(t)|e_j,\{0_k\}\rangle.\label{smct}
\end{equation}
According to the Floquet theorem, the evolution operator $\hat{\mathcal{U}}_j(t)$ is expanded as
\begin{equation}
\hat{\mathcal{U}}_j(t)=\sum_{\alpha}e^{-i\epsilon_\alpha t}|u_\alpha(t)\rangle\langle u_\alpha(0)|,\label{smut}
\end{equation}where $\epsilon_\alpha$ and $|u_\alpha(t)\rangle$ satisfy the Floquet eigen equation $[\hat{H}_j(t)-i\partial_t]|u_\alpha(t)\rangle=\epsilon_\alpha|u_\alpha(t)\rangle$ \cite{PhysRev.138.B979,PhysRevA.7.2203}.
Substituting Eq. \eqref{smut} into Eq. \eqref{smct}, we obtain
\begin{equation}
c(t)=\sum_{\alpha}e^{-i\epsilon_\alpha t}\langle e_j,\{0_k\}|u_\alpha(t)\rangle\langle u_\alpha(0)|e_j,\{0_k\}\rangle.
\end{equation}
Using the periodicity property $|u_\alpha(t)\rangle=|u_\alpha(t+T)\rangle$, we obtain $c(t)$ under the stroboscopic dynamics, i.e., $t = nT$ ($n\in\mathbb{Z}$), as
\begin{equation}
c(t)=\sum_{\alpha}e^{-i\epsilon_\alpha t}|\langle u_\alpha(0)| e_j,\{0_k\}\rangle|^2.\label{smst}
\end{equation}

The quasienergy spectrum of a periodically driven two-level system coupled to an environment constitutes a finite continuous quasienergy band and discrete quasienergy levels possibly formed, which we call Floquet bound states (FBSs). The contribution of the quasistationary states from the continuous quasienergy band to the long-time $c(nT)$ tends to zero due to the out-of-phase interference \cite{PhysRevA.91.052122,PhysRevA.102.060201}. Only the ones from FBSs survive. Then the long-time solution of Eq. (5) under the stroboscopic dynamics, i.e., $t=nT$, is
\begin{eqnarray}
\lim_{t\rightarrow\infty}c(t)=\sum_{l=1 }^MZ_{l}e^{-i\epsilon^\text{b} _{l}t}\label{smlotstd}
\end{eqnarray}
when $M$ FBSs are formed, where $Z_{l}=|\langle u^\text{b}_{l}(0)|e_j,\{0_{k}\}\rangle |^2 $ and $|u^\text{b}_{l}(t)\rangle$ is the $l$th FBS with quasienergy $\epsilon^\text{b}_{l}$.

Besides numerical calculation, the form of $\epsilon_{l}^\text{b}$ and $Z_{l}$ can be determined as follows. For simplicity, we consider only one FBS is formed and thus remove the label $l$. After making a Fourier transform $|u^\text{b}(t)\rangle=\sum_{n}e^{in\omega_{T}t}|\tilde{u}^\text{b}(n)\rangle$
and expanding $|\tilde{u}^\text{b}(n)\rangle=c_{n}|e_j,\{0_{\bf k}\}\rangle+\sum_{\bf{k}}d_{n}^{\bf{k}}|g_j,1_{\bf{k}}\rangle$, we recast the Floquet eigen equation into
\begin{equation}
\mathbf{Y}(\epsilon^\text{b})\cdot\mathbf{c}=\epsilon^\text{b}\mathbf{c},\label{eq:FBSs}
\end{equation}
where $\mathbf{c}\equiv \left(
                          \begin{array}{ccc}
                            \cdots & c_{n} & \cdots \\
                          \end{array}
                        \right)^T$ and $n\in\mathbb{Z}$
is the Fourier index. $\mathbf{Y}(\epsilon^\text{b})$ is a matrix with the matrix elements $[\mathbf{Y}(\epsilon^\text{b})]_{n,m}=\tilde{f}_{n,m}+[\omega_{0}+m\omega_{T}+\Sigma_{m}(\epsilon^\text{b})]\delta_{n,m}$, where $\Sigma_{m}(\epsilon^\text{b})=\sum_{\bf{k}}\frac{|g_{\bf{k}}|^{2}}{\epsilon^\text{b}-m\omega_{T}-\omega_{\bf{k}}}$ is the self-energy and collects all the environmental effects and $\tilde{f}_{n,m}=\frac{1}{T}\int_{0}^{T}f(t)e^{-i(n-m)\omega_{T}t}dt$ is the Fourier transform of control field $f(t)$. By solving Eq. \eqref{eq:FBSs}, the FBS quasienergy $\epsilon^\text{b}+m\omega_{T}$
in all Brillouin zones are obtained. Since Eq. \eqref{eq:FBSs} is translation invariant under displacement of $m\omega_{T}$, we only chose the first Brillouin zone formed by $\epsilon^\text{b}$.
The normalization condition $\langle u^\text{b}(t)|u^\text{b}(t)\rangle=1$ gives
\begin{equation}
\mathbf{c}^{\dagger}\cdot\mathbf{G}(\epsilon^\text{b})\cdot\mathbf{c}=1\label{eq:constrain}
\end{equation}
where $[\mathbf{G}(\epsilon^\text{b})]_{m,n}=[1-\partial_{\epsilon}\Sigma_{m}(\epsilon)]\delta_{m,n}\big|_{\epsilon\rightarrow\epsilon^\text{b}}$.
Combining Eqs. \eqref{eq:FBSs} and \eqref{eq:constrain}, $\epsilon^\text{b}$ and its corresponding $\mathbf{c}$ can be uniquely determined. Then $Z=|\sum_nc_n|^{2}$.

For the environment in the main text, we have
\begin{equation}
\Sigma_{m}(\epsilon^\text{b})=\frac{2g^{2}K[(\frac{4h}{\epsilon^\text{b}-m\omega_{T}})^{2}]}{\pi(\epsilon^\text{b}-m\omega_{T})},\label{eq:selfenergy}
\end{equation}
where $K(x)=\int_{0}^{\pi/2}[1-x\sin^{2}(\phi)]^{-1/2}d\phi$ is the complete elliptic integral of the first kind. By making the $\epsilon$-derivative to $\Sigma_{m}(\epsilon^\text{b})$, we have
\begin{equation}
[\mathbf{G}(\epsilon^\text{b})]_{m,n}=\{1+\frac{2g^{2}E[(\frac{4h}{\epsilon^\text{b}-m\omega_{T}})^{2}]}{\pi[(\epsilon^\text{b}-m\omega_{T})^{2}-(4h)^{2}]}\}\delta_{m,n},\label{eq:normalization}
\end{equation}
where $E(x)=\int_{0}^{\pi/2}[1-x\sin^{2}(\phi)]^{1/2}d\phi$ is the complete elliptic integral of the second kind.

\section{QFI in the presence of one FBS}
If one FBS is formed, i.e. $M=1$, then according to Eq. \eqref{smlotstd}, the steady-state $c(t)$, i.e., $t=nT$ reads
\begin{equation}
\lim_{t\rightarrow\infty}c(t)=Ze^{-i\epsilon^\text{b}t}.\label{smcstd}
\end{equation}
Substituting Eq. \eqref{smcstd} into Eq. \eqref{app:app10}, we readily obtain
\begin{eqnarray}
\lim_{t\rightarrow\infty}\mathcal{F}_{\omega_{0}}^{(2)} &=&\frac{N(\partial_{\omega_{0}}Z^{2})^{2}}{2 Z^{2}}\Big[{1-NZ^{2}(1-Z^{2})^{N-1}\over 1-Z^{2}}\nonumber\\&&-NZ^{2N-2}\Big],
\end{eqnarray}
which is time independent and tends to
\begin{equation}
\lim_{N,t\rightarrow\infty}\mathcal{F}_{\omega_{0}}^{(2)}={N(\partial_{\omega_0}Z^2)^2\over Z^2(1-Z^2)}.\label{smmf1}
\end{equation}
The steady-state form of $\rho_1(t)$ reads
\begin{equation}
\rho_1(t)=\frac{1}{2}\begin{pmatrix}Z^{2N} & Ze^{i\epsilon^\text{b}t}\\
Ze^{-i\epsilon^\text{b}t} & 1+(1-Z^{2})^{N}
\end{pmatrix}
\end{equation}
in the large-$t$ limit, we have its eigen solutions as
\begin{align}
\lambda_{\pm} & =\frac{s}{4}\Big[1\pm\sqrt{1-4Z^{2N}(1-Z^{2})^{N}/s^{2}}\Big],\\
|\lambda_{\pm}\rangle & =\left(
                               \begin{array}{cc}
                                 e^{i\epsilon^\text{b}N t}\sin\theta_{\pm} & \cos\theta_{\pm} \\
                               \end{array}
                             \right)^T,
\end{align}
where $s=1+Z^{2N}+(1-Z^{2})^{N}$ and
\begin{align}
\theta_{\pm} & =\arctan\frac{1}{2}\Big\{Z^{N}-Z^{-N}-(Z^{-1}-Z)^{N}\nonumber\\
&\pm\sqrt{s^2Z^{-2N}-4(1-Z^2)^{N}}\Big\}.\nonumber
\end{align}
The orthogonality between $|\lambda_{\pm}\rangle$ results in $\cos(\theta_{+}-\theta_{-})=0$. Keeping
the leading order of $t$ and $N$, we calculate from Eq. \eqref{app:app9}
\begin{equation}
\lim_{N,t\rightarrow\infty}\mathcal{F}_{\omega_{0}}^{(1)}\simeq y_{N}(Z^{2}) (Nt\partial_{\omega_{0}}\epsilon^\text{b})^{2},\label{smmf2}
\end{equation}
where
\begin{equation}
y_{N}(Z^{2})=\sum_{i=\pm}\lambda_{i}\sin^{2}(2\theta_{i})-\frac{16\lambda_{+}\lambda_{-}(\sin\theta_{+}\sin\theta_{-})^2}{\lambda_{+}+\lambda_{-}}. \label{smy}
\end{equation}
Further algebraic calculation reduces Eq. \eqref{smy} to
\begin{equation}
y_{N}(x)=\frac{2x^{N}}{x^{N}+(1-x)^{N}+1}.
\end{equation}
Combining Eqs. \eqref{smmf1} and \eqref{smmf2}, we finally obtain
\begin{equation}
\lim_{N,t\rightarrow\infty}\mathcal{F}_{\omega_{0}}(t)\simeq y_N(Z^2)(Nt\partial_{\omega_0}\epsilon^\text{b})^2+{N(\partial_{\omega_0}Z^2)^2\over Z^2(1-Z^2)}.
\end{equation}

\begin{figure}
\includegraphics[width=\columnwidth]{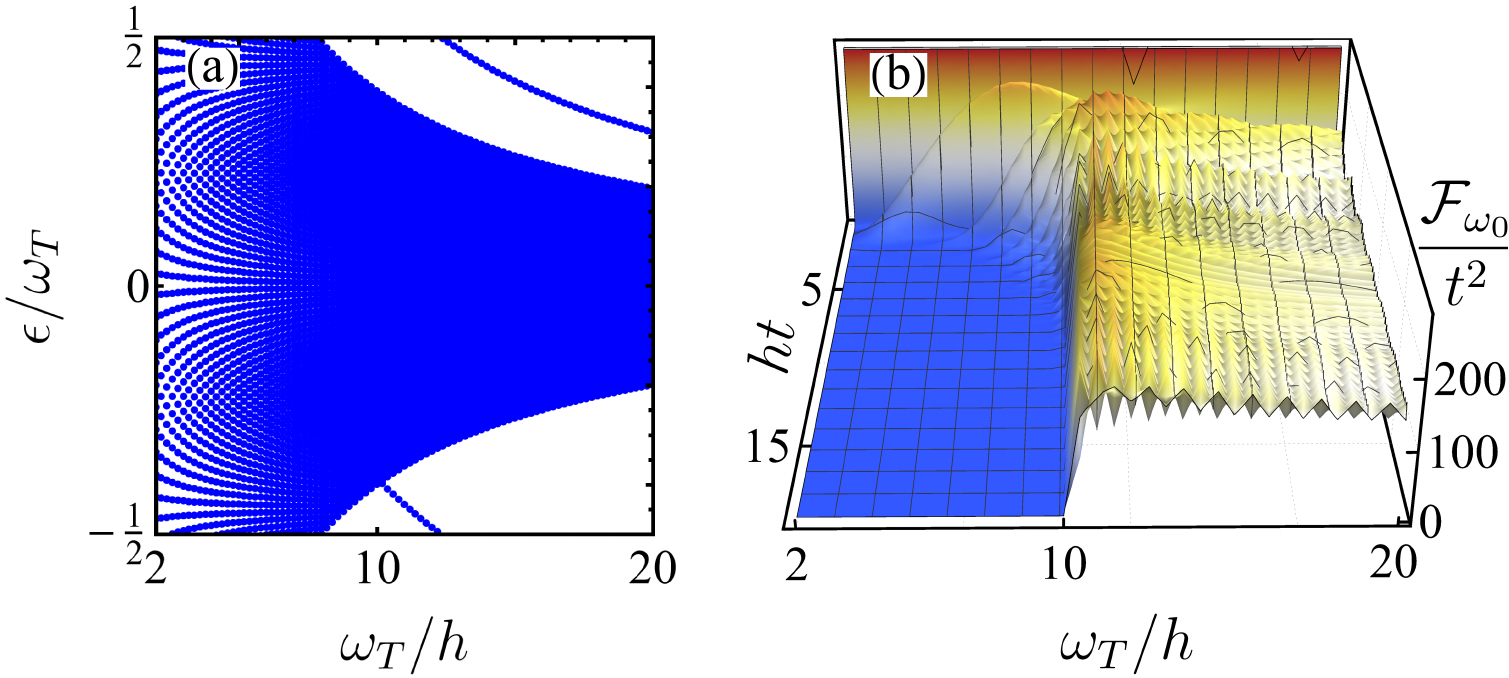}
\caption{(a) Quasienergy spectrum and (b) evolution of the QFI $\mathcal{F}_{\omega_0}/t^2$ in different driving frequency $\omega_T$. The ideal $t^2$ scaling of $\mathcal{F}_{\omega_0}$ is recovered in the regions where a FBS is formed. We used $A=10h$ and other parameters being the same as the ones in Fig. 2 of the main text. }\label{qesp}\end{figure}
\section{Quasienergy spectrum}
We here analyze the structure of the quasienergy spectrum of the total system formed by each periodically driven atom and its local environment. The environmental dispersion relation reads $\omega_{\bf k}=\omega_c-2h(\cos k_{x}+\cos k_{y})$, which forms a width $8h$ of the energy band centered at $\omega_c$. When a periodically driven two-level system coupled to this environment, the $8h$-width of the energy band is inherited by the quasienergy spectrum determined by the Floquet eigen equation  \cite{PhysRev.138.B979,PhysRevA.7.2203}
\begin{equation}
[\hat{H}_j(t)-i\partial_t]|u_\alpha(t)\rangle=\epsilon_\alpha|u_\alpha(t)\rangle,\label{smflqe}
\end{equation} where $\epsilon_\alpha$ and $|u_\alpha(t)\rangle=|u_\alpha(t+T)\rangle$ are called quasienergies and quasistationary states, respectively. The quasienergies $\epsilon_\alpha$ are periodic in a period $\omega_T$ because $e^{il\omega_Tt}|u_\alpha(t)\rangle$ is also the eigenstate of Eq. \eqref{smflqe} with eigenvalue $\epsilon_\alpha+l\omega_T$. One generally chooses them within $[-\omega_T/2, \omega_T/2]$ called the first Brillouin zone. The width of the quasienergy in the first Brillouin zone is $\omega_T$. After substituting the width $8h$ of the quasienergy band, a bandgap with finite width $\omega_T-8h$ can be present in the quasienergy spectrum only when $\omega_T > 8h$. In this parameter region, by properly changing $\omega_T$, the FBS may be formed in the band gap regime. As long as $\omega_T < 8h$, the quasienergy band fills up the full spectrum and thus there is no room for forming the FBS anymore. Therefore, a prerequisite for forming the FBS is the existence of the band gap in the quasienergy spectrum, which occurs when $\omega_T>8h$. 

We plot in Fig. \ref{qesp}(a) the quasienergy spectrum of the total system formed by the periodically driven two-level atom and the environment in different driving frequency $\omega_T$. It confirms our expectation that the band gap is present when $\omega_T>8h$. In this parameter region, with increasing $\omega_T$, the FBS is formed. The evolution of the QFI $\mathcal{F}_{\omega_0}$ in Fig. \ref{qesp}(b) reveals that in the region without the FBS, $\mathcal{F}_{\omega_0}$ decays to zero exclusively. When the FBS is formed, $\mathcal{F}_{\omega_0}/t^2$ tends to a non-zero constant with a persistent tiny-amplitude oscillation. Combining the result under the changing of the driving amplitude $A$ studied in the main text, we can see that, irrespective of which driving parameter is changed, the firm correspondence between the formation of the FBS and the recovery of the ideal $t^2$ scaling of the QFI $\mathcal{F}_{\omega_0}$ is valid. 

\bibliography{FM}